\documentclass{cambridge6A}
  \usepackage{natbib}
\usepackage[yyyymmdd,hhmmss]{datetime}
  \usepackage{rotating}
  \usepackage{floatpag}
  \rotfloatpagestyle{empty}
  
 \usepackage{amsmath}
  \usepackage{amsthm}
  \usepackage{amssymb}
  \usepackage{graphicx}

\usepackage{xcolor}
\usepackage[verbose,hypertexnames=false]{hyperref}
\hypersetup{colorlinks=false,allbordercolors=blue,pdfborderstyle={/S/U/W 1}}

\definecolor{IITred}{rgb}{0.5,0.05,0.05}
\definecolor{Dgreen}{RGB}{0,96,0}
\definecolor{Dcyan}{cmyk}{.96,0,0,.4}
\definecolor{IITblue}{rgb}{0.05,0.05,0.8}

    \newcommand{\ev}{\hbox{ eV}}

    \newcommand{\gev}{\hbox{ GeV}}

    \newcommand{\m}{\hbox{ m}}

\newcommand{\jpsi}{\ensuremath{J\!/\!\psi}}

\newcommand{\smgg}{\ensuremath{\mathrm{SU(3)_c} \otimes \mathrm{SU(2)_L} \otimes \mathrm{U(1)}_Y}}
\newcommand{\cgg}{\ensuremath{\mathrm{SU(3)_c}}}
\newcommand{\ewgg}{\ensuremath{\mathrm{SU(2)_L} \otimes \mathrm{U(1)}_Y}}
\newcommand{\ygg}{\ensuremath{\mathrm{U(1)}_Y}}
\newcommand{\wigg}{\ensuremath{\mathrm{SU(2)_L}}}
\newcommand{\emgg}{\ensuremath{\mathrm{U(1)}_{\mathrm{EM}}}}
\newcommand{\legg}{\ensuremath{\mathrm{SU(3)_c}  \otimes \mathrm{U(1)}_{\mathrm{EM}}}}

\newcommand{\shpp}[1]{$\langle \hbox{#1} \rangle$}
\renewcommand{\thesection}{\arabic{section}}
\begin{document}

  \title[Compiled on \today\ at \currenttime]
    {From the November Revolution\\ toward the Millennium}

  \author{Chris Quigg\\ Fermi National Accelerator Laboratory\\
  P.O. Box 500, Batavia, Illinois 60510 USA\\[3\baselineskip]
The 4th International Symposium  on the History of Particle Physics spanned two decades, from approximately 1980 to 2000. In this opening keynote lecture, I survey developments in particle physics from the prehistory of the \jpsi\ discovery to the early nineteen-eighties. I offer context for the contributions that follow in the symposium program. \\[24pt] \textsf{FERMILAB-PUB-26-0447-T}}
  \frontmatter
  \maketitle
 
 \let\cleardoublepage\clearpage 
  \mainmatter
  \chapter*{From the November Revolution toward the Millennium}
  
  \vspace*{-96pt}
  \begin{center}
  Chris Quigg \\
  Fermi National Accelerator Laboratory\\
  P.O. Box 500, Batavia, Illinois 60510 USA\\[12pt]
  Opening Keynote at the 4th International Symposium \\ on the History of Particle Physics at CERN, November 10--13, 2025\
  \end{center}
  \section{Prologue}
  On November 11, 1974, the dual announcement of the \jpsi$(3.1\gev)$ resonance catalyzed a phase transition in our understanding of the natural world.\footnote{A note on referencing: The period immediately preceding the 1980--2000 focus of this meeting was the subject of the third Symposium in this series, \cite{Hoddeson:1997hk}. I will not exhaustively cite contributions covered there. Earlier symposia are documented in \cite{Brown:1983,Brown:1989im}. I indicate contributions to this 4th Symposium as \shpp{Speaker}.} An experimental team working at Brookhaven National Laboratory's Alternating Gradient Synchrotron, \cite{PhysRevLett.33.1404}, observed a prominent narrow dielectron resonance in proton--beryllium scattering. A SLAC--Berkeley collaboration, \cite{PhysRevLett.33.1406}, announced a very sharp peak in the cross sections for $e^+e^- \to \hbox{hadrons, } e^+e^-, \hbox{ and possibly } \mu^+\mu^-$, at the SPEAR storage ring.  
  
 As thrilling as the announcement of the new resonance with exceptional properties was, the effect of that discovery was greatly amplified by ideas already in the air that began to meld together and give us a new picture of nature.\footnote{I developed this thesis in some detail in my lecture~\cite{CQSSI24} at SLAC, celebrating the fiftieth anniversary of the discovery.}   The cumulative effect is what we call the November Revolution. Few among us saw all of this coming.   Although physicists are said to be very self-confident, we don't always take our ideas as seriously as we later learn  that nature does. We do, after all, demand evidence!

Steven \cite{2545cd84-afae-3fe9-ae8a-0c6a9db8b3ac} has written  about the era of the Lamb shift, 
\begin{quote}``There is a huge apparent distance between the equations that theorists play with at their desks, and the practical reality of atomic spectra and collision processes. It takes a certain courage to bridge that gap and to realize that the products of thought and mathematics may actually have something to do with the real world. Of course, when a branch of science is well underway, there is continual give and take between  theory and experiment, and one gets used to the idea that the theory is about something real. Without the pressure of experimental data, the realization comes harder.
The great thing accomplished by the discovery of the Lamb shift was not so much that it forced us to change our physical theories, as that it forced us to take them seriously.''
\end{quote}

\begin{figure}[t]
\centerline{\includegraphics[width=0.5\textwidth]{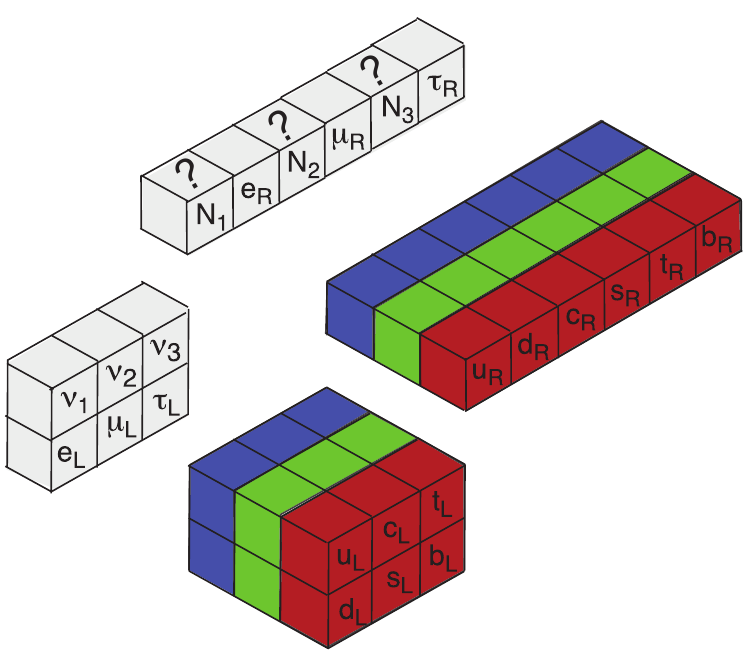}}
\caption{A schematic representation of the quarks and leptons, the fundamental constituents of the standard model of particle physics. The color-triplet quarks are painted red, green, and blue; the left-handed weak-isospin doublets are indicated by the stacked pairs. Interactions are governed by \smgg\ gauge symmetry, spontaneously broken to \legg.  \label{fig:SMcartoonA}}
\end{figure}

Much the same can be said of the period following the discovery of the \jpsi. Many of the ideas that we were exploring at the time turned out to fit together and to lead us to the more coherent understanding that we call the standard model of particle physics. 
At the most fundamental level that we have explored, our current conception is based on two sets of constituents, the quarks---particles with strong interactions---and the leptons---particles without strong interactions. These constituents interact by means of  forces given by gauge symmetries, as indicated in Figure~\ref{fig:SMcartoonA}. 

All of the basic constituents are spin-$\frac{1}{2}$ fermions. We idealize the quarks and leptons as having no discernible size. We observe them to be pointlike at our current resolution of less than  a billionth of a billionth of the human scale.\footnote{The CMS experiment,~\cite{CMS:2026ecv},  recently extended the limit toward $10^{-20}\m$.} Free quarks have never been seen, but free leptons are commonplace. 

The up, down, and strange quarks ($u,d,s$) were suggested already in the 1960s by the rich spectroscopy of hadrons---strongly interacting mesons and baryons---that were observed in bubble chambers and elsewhere. The notion of color was suggested so that three-quark baryon wave functions would respect the exclusion principle. We infer the left-handed quark and lepton doublets from the weak charged-current  transitions first observed in nuclear beta decay. The right-handed charged leptons carry weak hypercharge; the right-handed quarks carry both weak hypercharge and color. 

Quarks should come in pairs  to banish flavor-changing neutral currents. We could give indirect arguments that the $b$ quark, discovered in 1977, could not be a singleton, but its partner top quark was not discovered until 1995 \shpp{Grannis}. The matched quark and lepton doublets cancel anomalies---quantum corrections that do not respect the gauge symmetries---and so enable a consistent quantum theory of the electroweak interactions.

Even today, only left-handed neutrinos have been observed directly, so a 1980 version of Figure~\ref{fig:SMcartoonA} would have omitted the right-handed neutrinos. Because we have observed neutrino osciallations \shpp{Kajita, Smirnov}, I have included them tentatively  here, labeled $(N_1, N_2, N_3)$ with prominent question marks, reflecting the incomplete evidence that  right-handed neutrinos are required. Their existence and properties remain to be demonstrated.

The idea that symmetries dictate interactions,  or that function follows form, means that from the symmetries we observe in experiment, we might be able to derive the interactions that govern the relations between those fundamental particles. By trial and error and clever insight, we arrived at an \cgg\  color symmetry for the strong interaction, an \wigg\  of left-handed weak isospin inferred from charge-changing weak transitions such as beta decay, and a \ygg\ phase symmetry linked to weak hypercharge.

From today's perspective, when we already know a great deal of the answer, it may be tempting to imagine that in 1980 the entire standard-model paradigm had securely been established.
I want to assure you that for all of us who were working in particle physics at the time, whether in experiment, or in theory, or in accelerator science, or in instrumentation, it did not seem so. There was plenty to do, we did not know all the answers, we were certainly not sure of  the answers. 

Adding to the ferment, not every experimental result was either correct or correctly interpreted at first report. There was a time during the late 1970s when many people who took pleasure in imagining alternatives to what we now know as the canonical story could exercise their creative juices. It was a real playground for the model builders. There were results such as atomic parity violation experiments that didn't line up at all with the \ewgg\ electroweak theory. There were results in neutrino scattering, observations called the high-$y$ anomaly, super-trimuons, and such, that did not  survive, but gave us plenty to think about.

Indeed so much was happening that everyone could imagine being at the center of the action. Some people specialized immediately on one subject or another, others worked on many things at the same time, but it just seemed like a compelling frenzy of activity. It was {thrilling} to be part of it.\footnote{For first-person accounts of those revolutionary times, see \cite{Cahn:2023rwr}, Chapter 6.}

\section{Evolving the Theory of Strong Interactions \label{sec:smnow}}

A century ago, Emmy \cite{Noether:1918zz} proved  two far-reaching theorems. The first, which we celebrate in mechanics classes, establishes a  connection---in both directions---between global symmetries and conservation laws.  A second theorem, we now understand, shows how local symmetries imply interactions, and so points to the gauge principle that underlies much of our understanding of the fundamental interactions.

The first exemplar of the second theorem  was Hermann Weyl's construction of quantum electrodynamics from a $\mathrm{U(1)}$  symmetry applied to the phase of the quantum-mechanical wave function~\cite{Weyl:1918ib,Weyl:PNAS}. We  learn in our  first class on quantum mechanics that the absolute phase of the wave function $\Psi$ is not measurable. Change the convention by a certain fixed amount, $\Psi \to e^{ia}\Psi$, and the expectation values of Hermitian operators that correspond to observables, $\int\Psi^*\mathcal{O}\Psi$ \ldots are the same. 

But you can demand that this choice of the convention for phase not merely be a choice we make once and forever, but could be different in different times and places. Imposing that \emph{local} symmetry  then leads  to a theory with interactions---electrodynamics with the full content of Maxwell's equations. 

This strategy is so compelling, why not apply it somewhere else? The first internal quantum number, isospin, was invented in 1932, but no one jumped to apply the Noether--Weyl strategy to isospin symmetry, $\mathrm{SU(2)}_I$. Two decades later, \cite{Yang:1954ek} showed that the local non-Abelian isospin symmetry gave rise to massless isovector spin-one gauge bosons, $V^+, V^0, V^-$, that would mediate the force between nucleons and interact among themselves. The Yang--Mills construction is beautiful; it attracted enduring theoretical attention. But it doesn't describe the nucleon--nucleon interaction in the real world, which is dominated by the exchange of light pseudoscalar pions. 

By the early 1970s, it was considered great theoretical sport to  examine different field theories in order to prove (by exhaustion) that the phenomenologically successful quark--parton model lacked a rigorous basis. Theorists tested the hypothesis 
that no field theory existed in which partons or quarks behaved as nearly independent within hadrons but couldn't be liberated. The work of~\cite{Politzer:1973fx} and of~\cite{Gross:1973id} uncovered an exception, non-Abelian gauge theories. 

Thus was Yang--Mills theory  reborn for \cgg\ of color, rather than $\mathrm{SU(2)}_I$, as \emph{quantum chromodynamics} (QCD), with interactions mediated by massless color-octet gluons. A key aspect of the theory is that the strong coupling ``constant'' $\alpha_{\mathrm{s}}$, the analogue of the fine structure constant in quantum electrodynamics, depends logarithmically  on the scale or energy at which it is determined. In contrast to the behavior in electrodynamics, for which the coupling strength increases at short distances or high energies, $\alpha_{\mathrm{s}}$ decreases from a large value at low energies or long distances to progressively smaller values  at high energies or short distances. This behavior---denoted asymptotic freedom---raises the possibility of making reliable perturbative calculations for strong interactions.

Support for the reality of quarks accumulated from various scattering experiments, and from the properties of the \jpsi\ and  $\Upsilon$ families. By 1979, we had the first evidence for gluons as physical particles. Experiments at the PETRA storage ring (DESY/Hamburg),  produced three-jet specimens that we could classify as $e^+ e^-$ annihilations to quark $+$ antiquark $+$ gluon. 

Quantum Chromodynamics gave us the possibility of understanding how the parton model could have a theoretical basis, and how Bjorken scaling might be an approximate truth. It also made specific predictions for deviations from the scaling hypothesis. Those had not yet been verified in experiments in the 1970s. The quantitative running of the strong coupling toward smaller values at short distances was itself established during the period covered by this Symposium \shpp{Bethke}.   

The earliest  perturbative calculations of  strong-interaction phenomena were of undetermined precision. We were, I think, grateful to have an approach that gave semiquantitative insights, such as the inhibition of $\jpsi \to ggg$, which makes sense of the narrow width of \jpsi. Developing the techniques  not only to make these calculations, but also to give credible estimates of uncertainty was an extraordinary achievement \shpp{K.~Ellis}. 
Combining perturbative QCD with real-world experience led to parton-shower Monte Carlo simulations that would allow us to anticipate realistically both signals and backgrounds for our experiments \shpp{Webber}. These have become indispensable for modern collider experiments.

The comparison of the \jpsi\ and $\Upsilon$ families allowed us to begin to test the notion of flavor independence. According to QCD, the strong interaction should depend on the color coupling and not on the name of the quarks that are involved. A quantitative understanding of the strong-coupling regime of QCD, including calculations of the hadron spectrum and evidence for color confinement matured at a later time than the period of this meeting. Those problems  will, I trust, be treated in detail at the following Symposium. Even today, we lack a rigorous proof of confinement, so crucial foundational work remains to be done.

\section{Evolving the Electroweak Theory \label{sec:EWtheory}}
On the electroweak side, it was \cite{Glashow:1961tr} who pursued Fermi's analogy between the weak interactions and electromagnetism. In the 1930s when Fermi formulated his picture, it was by explicit reference to the picture suggested by radiation theory. He argued that the particles emitted in beta decay---the neutrino for one---were created in the interaction, not liberated from within the decaying particle. Glashow took this further and sought to make a relationship between electromagnetism and beta decay. He proposed the electroweak gauge symmetry as \ewgg. 

Glashow notes one little stumbling block: all the gauge bosons are massless, whereas the weak interaction was known to be short-range. Then he says, ``It is a stumbling block that we must put aside.'' So don't forget it, but put it aside and see where you can go. 
\cite{Weinberg:1967tq} and~\cite{Salam:1968rm} applied the lessons of spontaneous symmetry breaking developed by Goldstone, Nambu, Jona-Lasinio, Anderson, Englert, Brout, Higgs, Guralnik, Hagen, Kibble, and others, and created the electroweak theory that we all know.  Spontaneous symmetry breaking in our textbook electroweak theory is analogous to the Ginzburg-Landau description of superconductivity that gives mass to a photon inside a superconducting medium.

This approach gave the prediction of weak neutral currents, a massless photon, and massive $W^\pm$ and $Z^0$.  One of the first dramatic specimens of the new neutral-current interaction came from the Gargamelle bubble chamber experiment here at CERN: an event identified as the reaction $\bar{\nu}_\mu e \to \bar{\nu}_\mu e$  that cannot occur in the Fermi theory of weak interactions. 
There followed from this laboratory and others evidence for neutral-current interactions in neutrino--nucleon scattering, and a whole industry grew up to characterize the new phenomenon. 

The search for charm was animated because (as Glashow, Iliopoulos, and Maiani had argued) in order to have no flavor-changing neutral currents, it was not enough just to have the $(u,d,s)$ quarks posited in the 1960s. There had to be a charge $+\frac{2}{3}$ charm quark, for which we had mass estimates of about $1.5\gev$.  Moreover, the search for massive mediators of the weak interaction(s) now  had a plausible target for the mass. That prospect generated huge interest in the possibility of hadron colliders at sufficiently high energies, and eventually electron--positron $Z$-factories. 

\section{Quest for the Electroweak Gauge Bosons \label{sec:WandZ}}

At Fermilab, many of my experimental colleagues were fascinated by the prospect of making some sort of collisions that would make it possible to discover the $W$ and $Z$. Some proposed to collide protons from the 8-GeV Booster with the Main Ring at 400 GeV. Others imagined building new rings of one sort or another. In 1976, Alvin Tollestrup organized a Modest Colliding  Beams Workshop. This was the occasion at which we at Fermilab, and our community, received the notion that perhaps a proton-antiproton collider could become a practical reality. That spark colored the thinking of people very much. 

We analyzed what could be done in our Main Ring, which was comparable, in many ways, to CERN's Super Proton Synchrotron, but built in Bob Wilson's signature fashion, ``on a shoestring.'' 
We concluded that a Main Ring $\bar{p}p$ Collider would not  provide a secure future, that we would follow an orderly strategy. We would build a superconducting Energy Saver/Doubler first. We would do fixed-target collisions with 800-GeV protons and then, at the right moment, we would move on to colliding beams.

Leon Lederman presented this plan to the High Energy Physics Advisory Panel at a meeting on a weekend. The following week, there came a phone call from Washington saying, ``We can sell the collider, do that first.''  Over a weekend, my Fermilab colleagues made a design for the proton-antiproton collider, and in fact, that became the Tevatron collider.

My Paris colleague John Iliopoulos  likes to refer to the 1974 International Conference on High Energy Physics (London) as the last conference of the Dark Ages. The \emph{Proceedings} are very interesting reading because many observations just didn't seem to be making sense. Burt Richter gave a talk presenting the results from $e^+e^- \to \mathrm{hadrons}$ in which he concluded that nothing is working, Bjorken scaling is wrong, quarks are wrong, all of that. Leon Lederman announced that between 3 GeV and 10 GeV, there were no narrow resonances decaying into lepton pairs, also incorrect. \cite{Iliopoulos:1974gle} himself gave a visionary talk at that meeting in which he said, in effect, ``Suppose everything we've ever thought turns out to be true.'' And miraculously, that assertion is basically correct! 

By 1980, QCD and the electroweak theory had reached the stage of, ``We'd like to believe these. They're looking very plausible.'' But not all the evidence was in place. A lot of work  remained to be done.

Soon came the wonderful work here at CERN's S$p\bar{p}S$ Collider, notably the discovery of the   $W$ and $Z$ bosons by the UA1 and UA2 Collaborations \shpp{Jenni}, extended later at the Tevatron  \shpp{Shochet}.

This magnificent achievement attracted the attention of physicists, of course, but also of the United States government, and even of the  \cite{Europe3}. In an editorial entitled, ``Europe 3, U.S. not even Z-Zero,'' which looks as if it might have been drafted by the President's Science Advisor, the ``newspaper of record'' argued that the U.S. had better get its act together, otherwise American particle physics would be obliterated by the progress in Europe.  This was the \emph{Zeitgeist} in which the Superconducting Super Collider was  conceived.

Further support for the implications of the  \ewgg\ electroweak theory would come from HERA, the $e^\mp p$ collider at DESY. At large values of the momentum transfer $Q^2$, the interaction rates for charged-current and neutral-current reactions are comparable.

If the electroweak symmetry is hidden, how can you show that it is present? By examining the interplay among different diagrams (electron exchange, photon exchange, $Z$ exchange) that contribute to the reaction $e^+e^- \to W^+W^-$,  you can witness the power of gauge symmetry. Each diagram by itself would give an unacceptable growth with energy, but the well-behaved cross section measured at LEP, the Large Electron--Positron Collider, validates the electroweak theory prediction. 

Still unresolved in 1980 was the notion of precisely how the electroweak symmetry was spontaneously broken or hidden. It's manifest that \ewgg\ is not a symmetry of the everyday world, whereas the phase symmetry of electromagnetism, \emgg, is. Whether the agent of electroweak symmetry breaking was  a single Higgs field, or  some dynamical process, or something else, we did not know.

\section{The Higgs Sector \label{sec:Higgs}}
From the late 1970s, we focused on the TeV scale ($10^{12}\ev$) as the place where we would learn the nature of electroweak symmetry breaking \shpp{J.~Ellis}. That target formed a great deal of motivation for the Superconducting Super Collider and the Large Hadron Collider, for the parameters of those machines, and for how we conceived of the detectors that would be built to exploit them. Our confidence was rewarded by the discovery in 2012 of a putative Higgs boson, $H(125\gev)$.\footnote{The search and discovery will be a highlight of  the next Symposium. For summaries of what was learned in the  decade after the discovery, see \cite{ATLAS:2022vkf,CMS:2022dwd}.} The decays of this unstable, neutral scalar particle into $W^+W^-$ and $ZZ$ implicate $H$ as an agent of electroweak symmetry breaking.

Evidence is developing today as if $H(125)$ were indeed the textbook Higgs boson, but the case is not yet closed. We have not yet measured the $H$ self-interactions to determine the shape of the Higgs potential.

One of the astute features about the Weinberg--Salam construction was not only that the Higgs field would hide the electroweak symmetry, giving mass to the $W^\pm$ and $Z^0$ while leaving the photon massless, but that Yukawa couplings between the Higgs field and the fermions  would generate masses for the quarks and charged leptons. Experiments at the Large Hadron Collider are finding that the Higgs couplings are indeed proportional to the masses of the fermions. That pattern doesn't exclude other explanations but it makes them seem a little superfluous. 

That is a wonderfully satisfying result. However, even as it validates a key feature of the electroweak theory, it implicates  physics beyond the standard model. There is within the standard model, no calculation you can do or even imagine that tells you what the values of those fermion masses should be.  There have been many ideas and much creative work. I think we never got more out than we put in. So that is an open question and a very important one, even today. 

The gauge interactions all embody universality. It is the gauge charge that matters and not the name of a particle. The Higgs-fermion interactions are decidedly non-universal and we don't understand what that means. Tini Veltman, who had a way with words, liked to grumble, ``The Higgs boson knows something we don't know,'' because spontaneous symmetry breaking determines not only the fermion masses, but also the mixing angles among the quarks and the complex phase that gives rise to hadronic CP violation. 

Mysteries remain to be resolved, even if this is the right track for our theory of the weak and electromagnetic interactions. In a way that we perhaps emphasize too little in everyday practice, flavor physics is intimately connected with Higgs physics. We'll hear in this meeting about charm and bottom spectroscopy \shpp{Cassel}, setting the scene for testing the quark-mixing matrix  (Cabibbo--Kobayashi--Maskawa) paradigm as the origin of CP violation. What might we learn by considering the flavor properties of quarks and leptons together?

\begin{figure}
\centerline{\includegraphics[width=0.40\textheight]{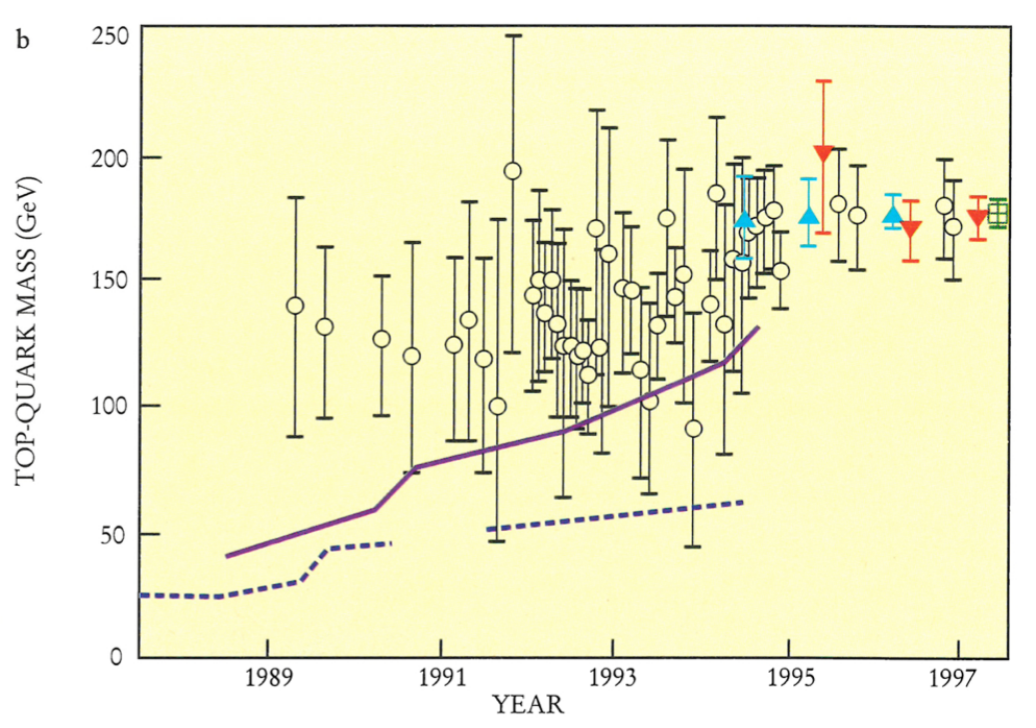}}
\caption{Evidence for the top mass over time. Rising lower bounds  inferred from direct searches are shown for $\bar{p}p$ colliders (solid line),assuming that standard decay modes dominate, and for $e^+e^-$ colliders (dashed lines). Open circles are indirect determinations  from fits to electroweak 
measurements. Colored triangles are direct measurements by CDF (\textcolor{cyan}{$\blacktriangle$}) and D0 (\textcolor{red}{$\blacktriangledown$}) at the time of initial evidence, discovery claim, and 1997. The green box indicates a world average of direct measurements.  
From~\cite{Quigg:1997uh}.\label{fig:tts}} 
\end{figure}

Quantum corrections to the lowest-order predictions of the \ewgg\ electroweak theory have a marked dependence on the mass of the top quark. Before top was observed, it was natural to take the accumulating data from the Stanford Linear Collider  \shpp{Swartz},  LEP \shpp{Pepe-Altarelli}, and other sources, and use them to infer what the mass of the top quark might be. I have collected many of these early determinations in Figure~\ref{fig:tts}.
From the earliest times, around 1989, these indirect analyses suggested that the top-quark mass might be much larger than the 4.2-GeV mass of the $b$ quark. By the time that the top quark was discovered in 1995, a consensus of the indirect inferences pointed to a range  between 150 and $200\gev$. The agreement with the observed mass of $175\gev$ was an early success for the electroweak theory as a quantum field theory. Ultimately, the accord between theory and experiment reached part-per-mille precision for many observables.

\section{A Wider View \label{sec:BSM}}
To conclude, I want to mention a few topics that emerged in the years preceding the period covered in this symposium.
\subsection{(Grand) Unification \label{subsec:GUT}}
One appealing extension is the idea of a  unified theory of strong, weak, and electromagnetic interactions \shpp{Dimopoulos}. This idea builds on the (partial) unification of the weak and electromagnetic interactions embodied in the \ewgg\ electroweak theory. That is but a partial unification because it entails two independent couplings.

Look back to Figure~\ref{fig:SMcartoonA}. It is tempting to slide the matching families of quarks and  families of leptons  together and make extended families of quarks and leptons.  We might think of lepton number as a fourth color. Once we join quarks and leptons in extended families, we open the possibility of quark--lepton transitions and  new phenomena such as proton decay.

\begin{figure}
\centerline{\includegraphics[height=0.25\textheight]{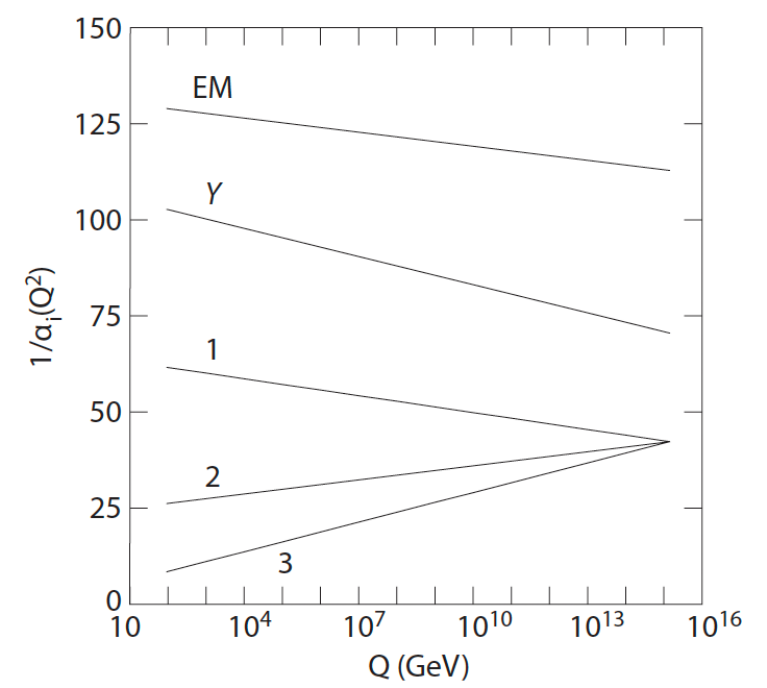}\qquad\includegraphics[height=0.25\textheight]{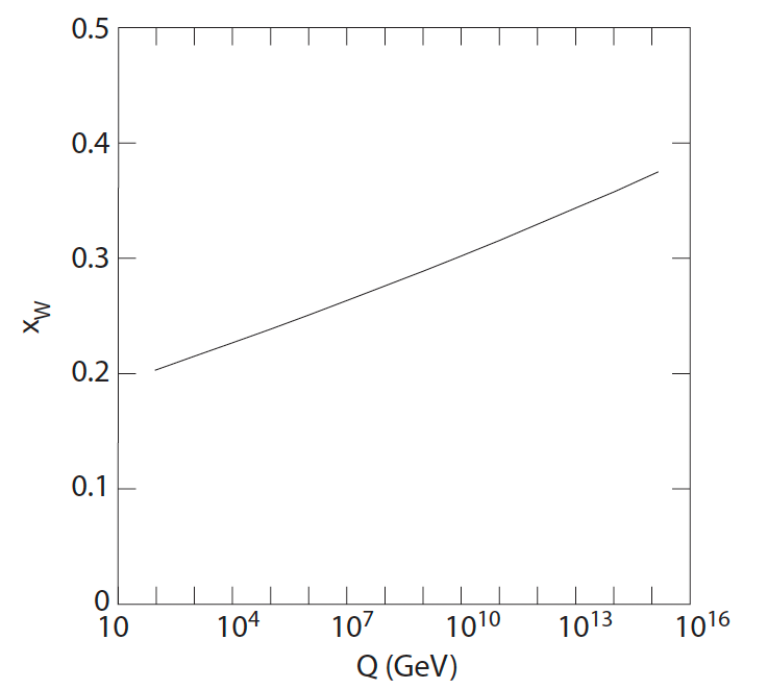}}
\caption{Left panel: Idealized evolution of running coupling constants in leading logarithmic approximation
is the SU(5) model. Three fermion generations are assumed. Right panel: Evolution of the weak mixing parameter $x_W =\sin^2{\theta_W} = \alpha_{\mathrm{EM}}/\alpha_2$  in the idealized SU(5) model. \label{fig:su5}}
\end{figure}

Unified theories also make quantitative predictions for our low-energy observables.
In the left panel of Figure~\ref{fig:su5} I show in a schematic picture of SU(5) unification, the evolution with energy $Q$ of the different couplings, $\alpha_1^{-1}, \alpha_2^{-1}, \alpha_3^{-1}, \alpha_Y^{-1}$ and, as a derived quantity, $\alpha^{-1}_{\mathrm{EM}}$. There is the quite fascinating notion that the symmetries of a unified theory would predict the weak mixing parameter, $\sin^2{\theta_W}$. At the unification scale, where everything is beautiful, the SU(5) theory predicts that $\sin^2{\theta_W} =3/8$. That does not agree with low-energy experiments. 
But, through work done by~\cite{Georgi:1974yf}, we discovered that if we evolved that prediction down to low scales where we do experiments, we found a number around 0.2--0.21. That  was just what seemed to be indicated by experiments at that time. By going away from our everyday world to a more perfect world at a unification scale, we could get insights that would then enlighten us down here. That is a very powerful notion!
\subsection{Supersymmetry \label{subsec:SUSY}}
If you took a vote among theorists, supersymmetry \shpp{Barbieri}, which links fermions and bosons,  was the most compelling extension of the standard model. The first motivation for taking it seriously is, why would nature neglect such a beautiful idea? Within supersymmetry, we could find a reason for the electroweak symmetry to be spontaneously broken. That was kind of seductive. The lightest supersymmetric particle might be a stable particle that could serve as dark matter. A spectrum of superpartners would improve the coupling-constant unification from that schematic picture that  we have just reviewed. Experiments at LEP and elsewhere encourage that possibility, but direct searches for superpartners at LEP \shpp{Rembser} and elsewhere have been unavailing.

String theory blossomed in the 1980s and remains a thriving element of theoretical research \shpp{Lerche}.
\subsection{Cosmic Physics \label{subsec:cosmic}}
This period leading to 1980 also saw the growth of the cosmic connection with particle physics \shpp{Kolb, de Swart}. Unified theories suggested, along with the instability of the proton, that the Sakharov conditions to explain the baryon asymmetry of the universe were at hand. There was accumulating evidence for dark matter and the existence of plausible candidates, including axions \shpp{Turner, Quinn}. Inflation \shpp{Guth} gave us a graceful hypothesis for the state of the universe.  Steven~\cite{Weinberg1977}'s \emph{The First Three Minutes}  encouraged others that the history of the universe was an essential area of study. 
\section{What we will hear this week \label{sec:TK}}
This week, we'll hear many stories of ferment, promise, invention, discovery, heartbreak, and triumph. We will see that the decades 1980--2000 are not a period of mere consolidation. If that were so, why would we have this meeting? We will be hearing about magnificent instruments---accelerators, colliders, and detectors, not to mention the integration of large-scale computing and the influence of the World Wide Web \shpp{Hoogland, White}. 

We will see examples of our science in service to society and of science in dialogue with society, including educational initiatives. And we will hear something of how projects succeeded and failed---decidedly human stories.

We will hear from people who have made our instruments and have made inventions that make those instruments possible. These beautiful devices often start with a very modest step. If you go  to the Rijksmuseum Boerhaave in Leiden,   you can see the first superconducting coil made of lead wire cooled below 7 kelvins. It fits in your hand;  it reached a staggering field of 600 gauss,  a first step along the path that led to the LHC magnets and beyond.

\section*{Acknowledgment}
I thank the Organizers, our CERN hosts, and all participants for a lively, stimulating, and enriching Symposium. 

This work was produced by Fermi Forward Discovery Group, LLC, under Contract No. 89243024CSC000002 with the U.S. Department of Energy, Office of Science, Office of High Energy Physics. Publisher acknowledges the U.S. Government license to provide public access under the DOE Public Access Plan.

\let\cleardoublepage\clearpage




  \backmatter


 \renewcommand{\refname}{Bibliography}

  \bibliography{CQHSPP}\label{refs}

  \bibliographystyle{cambridgeauthordate}

\end{document}